\documentclass[aps,preprint]{revtex4}%
\usepackage{amsfonts}
\usepackage{amsmath}
\usepackage{amssymb}
\usepackage{graphicx}
\renewcommand{\narrowtext}{\begin{multicols}{2} \global\columnwidth20.5pc}

\newcommand{\Fig}[1]{Fig. \ref{#1}}
\def\be{\begin{eqnarray}}
\def\ee{\end{eqnarray}}

\newcommand{\ra}{\rightarrow}

\begin{document}
\draft

\title{On the Dichotomy between the Nodal and Antinodal Excitations in High-temperature
Superconductors}
\author{Henry Fu and Dung-Hai Lee}
\affiliation{Department of Physics,University of California at
Berkeley, Berkeley, CA 94720, USA\\
Material Science Division, Lawrence Berkeley National
Laboratory,Berkeley, CA 94720, USA.}

\date{\today}
\begin{abstract}
Angle-resolved photoemission data on optimally- and under-doped high
temperature superconductors reveal a dichotomy between the nodal and
antinodal electronic excitations. In this paper we propose an
explanation of this unusual phenomenon by employing the coupling
between the quasiparticle and the commensurate/incommensurate
magnetic
excitations.
\end{abstract}

\maketitle
\parindent 10pt

\begin{figure}
\centering
\vskip -3.5 in
\includegraphics[width=4in]{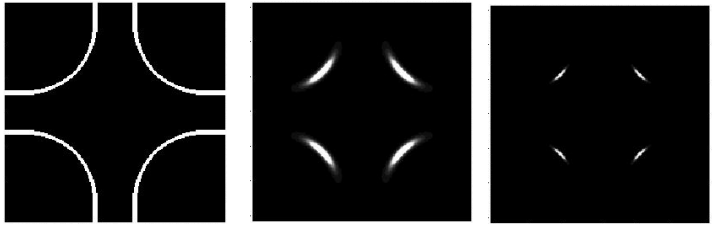}
\caption{The Bogoliubov quasiparticle weight $z$ along the normal
state Fermi surface as observed by ARPES. The brightness is proportional to
the magnitude of $z$. The doping decreases from the left to right.
 \label{dich}}
\end{figure}

Angle-Resolved Photoemission Spectroscopy (ARPES)\cite{zxreview} has
made important contributions to the understanding of
high-temperature superconductors.  The information revealed by this
technique has pointed to an unusual dichotomy\cite{zhou} between the
nodal and antinodal electronic excitations. In particular, as the
Mott insulating state at low doping is approached, the quasiparticle
weight vanishes on part of the Fermi surface (the antinodal region)
while it remains finite on the rest (the nodal region). This is
schematically illustrated in \Fig{dich}.  We refer to this strong
momentum-dependence of the quasiparticle weight as the dichotomy
between the nodal and the antinodal exciations.

In the rest of the paper we first describe the experimental
evidence from ARPES leading to this characterization of the
nodal/anti-nodal dichotomy. Following that we propose a mechanism
for the origin of this
phenomenon.\\

{\noindent{\bf{Nodal-antinodal dichotomy in ARPES}}}\\

\begin{figure}
\centering\vskip 0 cm\hskip -1.5 cm
\includegraphics[width=3in, angle=-90]{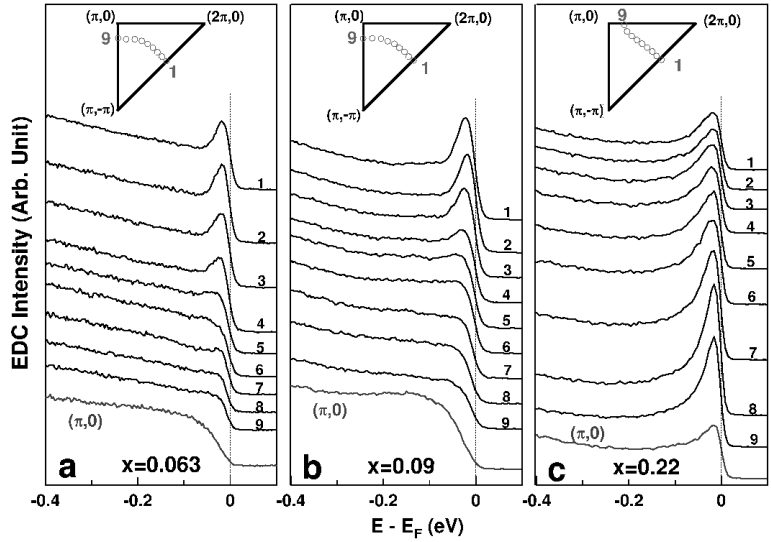}\vskip 1cm
\caption{The nodal (1) to antinodal (9) ARPES spectra for
$La_{2-x}Sr_xCuO_4$ at doping $x=0.063, 0.09, 0.22$.  From
Zhou {\it et al} \cite{zhou}.\label{lsco}}
\end{figure}

\Fig{lsco} illustrates the node $\ra$ antinode ARPES spectra for
$La_{2-x}Sr_xCuO_4$ at a fixed temperature $\sim 20K$ of Zhou {\it
et. al.}\cite{zhou}. The doping levels for the three panels are
$0.063, 0.09$ and $0.22$ from left to right. For the $x=0.22$ 
(overdoped) sample a quasiparticle peak is observed at all points on
the Fermi surface. In contrast, at $x=0.063$ the quasiparticle peak
only exists within a fixed angular range around the node. Similar
nodal quasiparticle peaks are observed in even $3\%$ doped
samples\cite{3pec}.

\begin{figure}
\centering
\includegraphics[width=2in,angle=0]{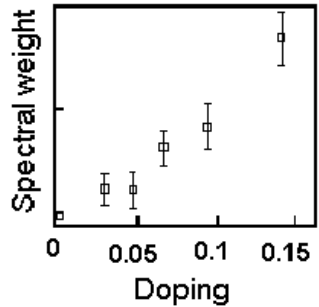}
\caption{The spectral weight of nodal quasiparticle
peak as a function of doping. From K. Shen {\it et al}\cite{kshen}.
\label{weight}}
\end{figure}

It should be noted that although the nodal quasiparticle peak exists
for all doping, its spectral weight does diminish as $x\ra 0$ (see
\Fig{weight})\cite{kshen}. This diminishing of the quasiparticle
weight is well described by a class of theories based on using the
Gutzwiller projected wavefunction to described the strongly
correlated electronic states\cite{xw}. However, these theories do not
explain the
interesting fact that while nodal excitations are well-defined
quasiparticles, the antinodal excitations are completely decoherent.
\\
 
{\noindent{{\bf{A mechanism for the antinodal decoherence}}}
\\

Here we propose a mechanism for the antinodal decoherence that
focuses on the role of magnetic excitations and their coupling to
the antinodal quasiparticles.
Before we begin, we present
two experimental clues to the origin of antinodal decoherence: the absence of
a large leading edge gap in ARPES measurements of the antinodes,
 and the existence of low-energy spin excitations.
\\

First, a close-up of the leading edge behavior of the ARPES spectra
near the antinode (enclosed by the box in \Fig{edge}) for the
$6.3\%$ doped LSCO\cite{zhou} is illustrated in \Fig{edge}.  A close
inspection shows that the set back of these leading edges is only
about $10 meV$. For doping as low as $x=0.063$ such a small gap is
very surprising, because from other measurements, e.g. NMR, the
pseudogap should increase with underdoping\cite{tallon}. Hence at
$x=0.063$ one would expect a much larger gap. This leading edge
behavior tells us that there are low-energy excitations with the
quantum number of a photohole which are not coherent quasiparticles.
\\
\begin{figure}
\centering
\includegraphics[width=2in,angle=0]{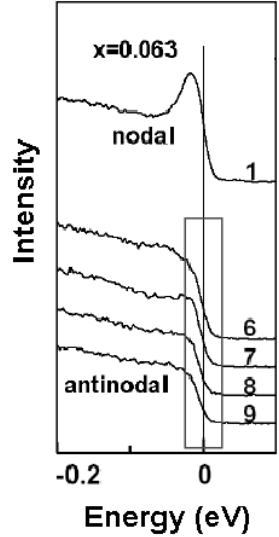}
\caption{The set back of the leading edge near the antinode
(enclosed by red box) is only $\approx 10 meV$. The spectra are taken at momenta labelled as in \Fig{lsco}a.
From Zhou {\it et al} \cite{zhou} .\label{edge}}
\end{figure}


Secondly, it has been well established that in LSCO there exist low energy
spin excitations in the neighborhood of momentum
$(\pi,\pi)$\cite{yamada}. For example, at $6\%$ doping, inelastic
neutron scattering demonstrates enhanced spectral weight around
$(\pi\pm\delta,\pi)$ and $(\pi,\pi\pm\delta)$ for energies as low
as $2$ meV (see \Fig{yamada}). In the following we propose that
the electronic excitations contributing to the leading edge
spectral weight are continuum excitations made up of low energy
spin excitations and quasiparticles near the nodes.
\\
\begin{figure}
\centering
\includegraphics[width=2in,angle=0]{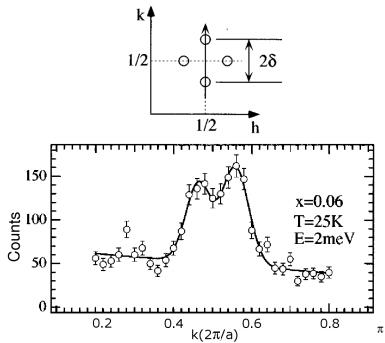}
\caption{The existence of low energy commensurate/incommensurate
magnetic excitations in 6\% doped LSCO. From Yamada {\it et
al} \cite{yamada}.\label{yamada}}
\end{figure}


{\bf A mechanism for the antinodal decoherence}\\

 For momenta
equal to those of the nodes (dot A  of \Fig{mech}), the lowest
energy excitation consistent with the quantum number of a photohole
is the zero-energy quasiparticle. As the momentum moves toward the
antinode, the quasiparticle gap increases. It is possible that that
at an intermediate momentum between the node and the antinode, the
lowest energy excitation ceases to be a quasiparticle. For example,
for momentum at the Brillouin zone face (indicated by dot B in
\Fig{mech}), a multi-particle excitation with energy lower than the
quasiparticle can exist. We propose that this type of multi-particle
excitation consists of a quasiparticle with momentum close to the
node (dot C in \Fig{mech}) and an incommensurate spin excitation
with momentum indicated by the the arrow. Such multi-particle
excitations contribute to the leading edge of the ARPES spectrum
near the antinodes. Since as a function of excitation energy, the
gapped quasiparticle states are preceded by this multi-particle
continuum, they can no longer be coherent. This is because energy
conservation allows them to decay into the multiparticle states.

Clearly, in order for the above mechanism to work,  the spin
excitation must cost sufficiently low energy. If this requirement
is not met, antinodal quasiparticle peaks will be exhibited and
the leading edge will be determined by the quasiparticle gap.
Under such condition the nodal-antinodal dichotomy is absent. We
expect this to happen when the doping is sufficiently high.\\

\begin{figure}
\centering
\includegraphics[width=2in,angle=0]{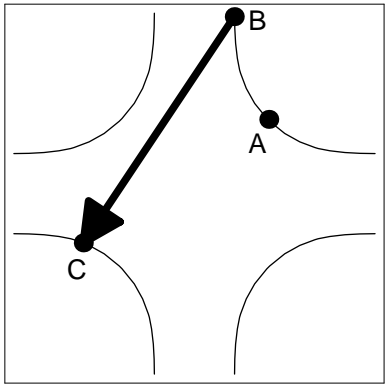}
\caption{Schematic illustration of the mechanism of antinodal
decoherence.\label{mech}}
\end{figure}

{\bf A renormalization group perspective}\\

 Starting from the
overdoped side, which is widely believed to be a Fermi liquid, we
expect that decreasing doping introduces residual quasiparticle
interactions. For doping that is not too low, the effects of these
residual interactions can be analyzed in a perturbative
renormalization group (RG) approach. This point of view has been
adapted by Rice and coworkers\cite{rice}, and has been shown to
capture much of the cuprate phenomenology in the appropriate doping
range. Recently Fu {\it et al} generalized this approach to include
the quasiparticle-phonon interaction\cite{fu}.

In the following we present the results of pure electronic
quasiparticle scattering using a realistic Fermi surface. The
qualitative nature of our results remain unchanged as long as the
residual quasiparticle interaction is not too weak and the Fermi
surface shows a nested antinodal region. The RG analysis is
performed numerically by discretizing the first Brillouin zone into
32 patches. All one-loop diagrams are included. In
\Fig{RGsupport}(b) the renormalized scattering amplitude is plotted
as a function of the two incoming momenta $\vec k_1$ (vertical axis)
and $\vec k_2$ (horizontal axis) while $\vec k_3$  is fixed at the
position marked by dot number two in \Fig{RGsupport}(c) and
\Fig{RGsupport}(d). The scattering processes that are dominantly
enhanced by the RG flow are those enclosed in the boxes labelled ``A''.  In
these vertical boxes there is a nearly constant momentum transfer
$\vec k_2-\vec k_3$ in the spin exchange channel. As a result we
identify them as being responsible for the spin fluctuations with
momenta near $(\pi,\pi)$, including the ``incommensurate'' momenta
such as $(\pi\pm\delta,\pi)$ and $(\pi, \pi\pm\delta)$.
Interestingly, this class of scattering processes involves primarily
the antinodal quasiparticle states on the Fermi surface (see
\Fig{RGsupport}(c)). The fact that only states on the Fermi surface
are involved in these scattering processes implies that the
corresponding spin fluctuations have low-energy. In contrast, all
RG-enhanced scattering processes involving only nodal quasiparticles
have states off of the Fermi surface. As a result they lead to
higher energy spin fluctuations (see \Fig{RGsupport}(d)). This is
consistent with the proposal that this type of quasiparticle
scattering is responsible for the 41 meV neutron resonance at
$(\pi,\pi)$.\cite{BrinckmannLee} Since these scattering processes
must involve high-energy quasiparticles, they do not lead to
decoherence of the nodal quasiparticles.

\begin{figure}
\centering
\includegraphics[width=6in,angle=0]{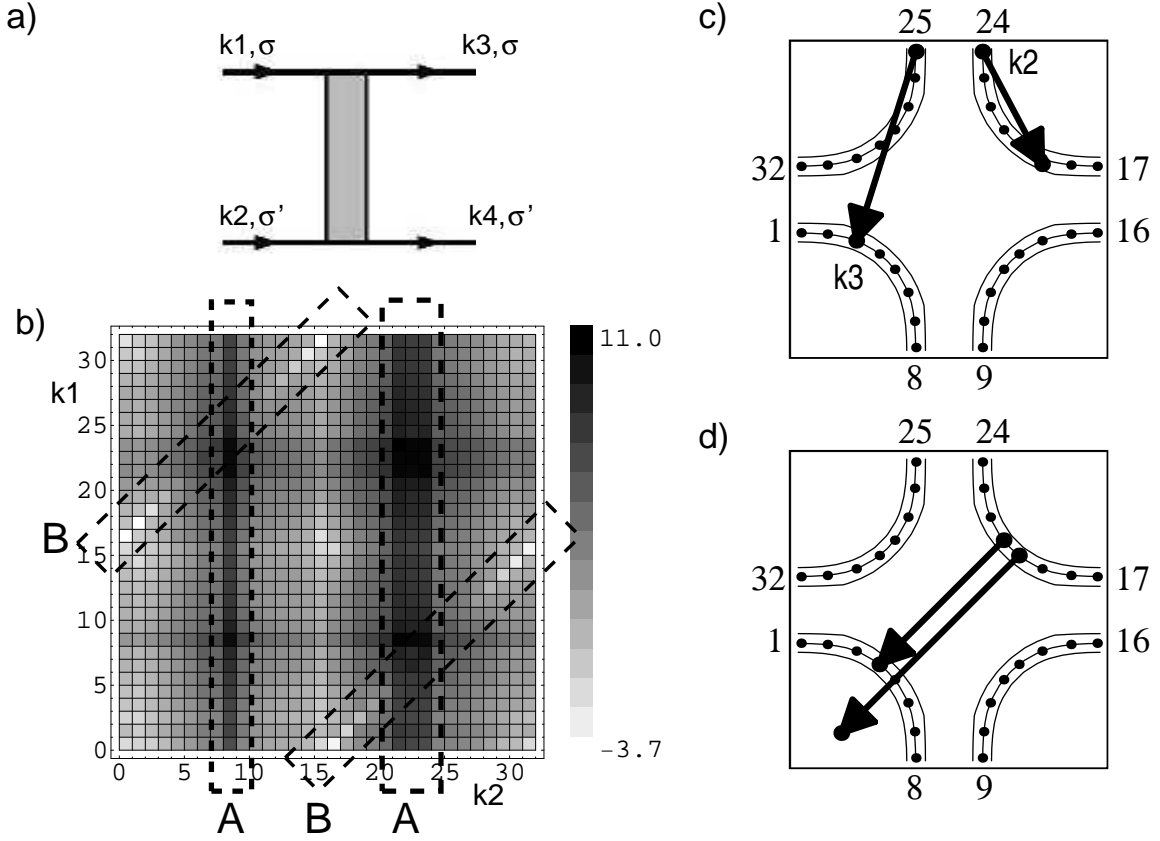}
\caption{The renormalized quasiparticle scattering.  a) The
quasiparticle scattering vertex. Spin is conserved along solid
lines.  Each of $\vec k_1$, $\vec k_2$, $\vec k_3$, and $\vec k_4$
lies in one of the 32 radial patches of the discretized Brillouin
zone. The centers of the intersection between the fermi surface and
the patches are shown as black dots in parts (c) and (d). The
patches are indexed counterclockwise from 1 to 32 as shown in the
Figure. b) The renormalized quasiparticle scattering amplitudes
plotted as a function of $\vec k_1$ and $\vec k_2$ when $\vec k_3$
is fixed at the second dot.  The strongest scattering amplitudes are
in the boxes labelled ``A'' . 
Common among all such strong scattering processes is
the momentum transfer $\vec k_2-\vec k_3\approx (\pi,\pi)$, i.e.,
the momentum transfer in the spin spin-exchange channel. In
addition, all such scattering processes involve electronic
excitations in the antinodal region. Aside from the strongest
magnetic scatterings, the diagonal boxes labelled ``B'' correspond to
attractive scattering in the d-wave cooper pair channel. c) An
example of the scattering processes that lead to low energy magnetic
fluctuations at momentum $(\pi-\delta,\pi)$. Note that these
scattering processes involve antinodal quasiparticle states.
d) An example of the scattering processes that lead to higher energy
spin fluctuation at momentum $(\pi,\pi)$. Note that these processes
involve quasiparticle states in the nodal direction only.
\label{RGsupport}}
\end{figure}

\begin{figure}
\centering
\includegraphics[width=5in,angle=0]{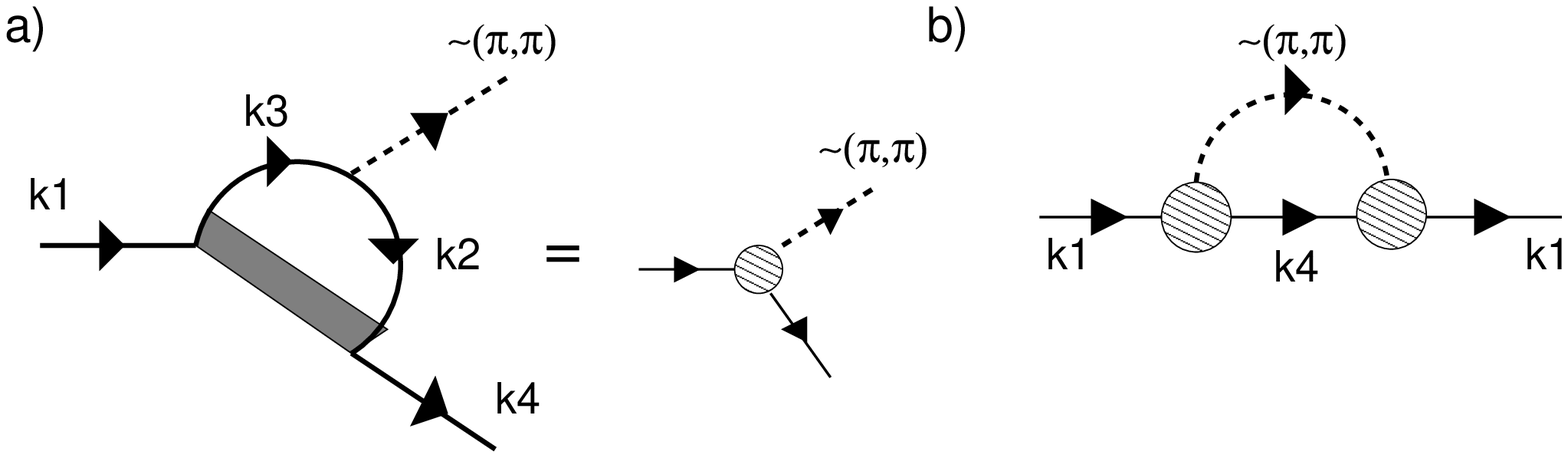}
\caption{ a) The interaction of electrons with spin excitations
using the
  strongly renormalized electronic couplings from \Fig{RGsupport} as
  vertices.  The dashed line is an outgoing low-energy magnetic excitation.
  b) Contribution to the single-particle spectral function which is enhanced
  by the strongly renormalized couplings through the vertex of part (a).  The
  internal loop corresponds to the multiparticle excitation discussed in the
  text. \label{vertex}}
\end{figure}

Are the above RG results consistent with the antinodal decoherence
mechanism we proposed earlier?  Consider the strongest low energy
quasiparticle scattering processes such as \Fig{RGsupport}c. Note
that while momentum $\vec k_2$ lies on the zone boundary, momentum
$\vec k_3$ lies closer to the nodal region. This is similar to the
quasiparticle component of the multiparticle excitation in
\Fig{mech}. Indeed, this scattering process contributes to the
vertex describing the scattering of a antinodal excitation into a
near nodal quasiparticle with the emission/absortion of a low energy
commensurate/incommensurate magnetic excitation, as shown in
\Fig{vertex}(a) and \Fig{vertex}(b). This is precisely the process
we invoke in the antinodal decoherence mechanism !\\

{\bf Single-hole ARPES and spin waves}\\

 The ARPES result of
insulating cuprates such as $Sr_2CuO_2Cl_2$\cite{zxreview,scoc} has
attracted much discussion and attention in the past. For such
compounds, the sharp coherent quasiparticle peak (near momenta
$(\pm\pi/2,\pm\pi/2)$) is replaced by an incoherent broad hump. The
hump has an {\it isotropic} dispersion in the shape of a cone with
its tip at momentum $(\pm\pi/2,\pm\pi/2)$. Interestingly, the slope
of the dispersion is basically the same as the spin wave velocity in
the antiferromagnet.\cite{greven}
\\

This intriguing result has stimulated many theoretical works
proposing that the cone-like dispersion is due to the spinon of a
spin liquid (which is predicted to have an isotropic, cone-like,
dispersion). In view of the decoherence mechanism proposed earlier,
here we would like to suggest an alternative, more mundane,
scenario. We propose that the broad dispersing feature seen in ARPES
actually arises from the multi-particle states composed of a
quasiparticle at momenta $(\pm \pi/2,\pm \pi/2)$ and a spin wave (see Fig. 9c).
The isotropic cone is precisely the spin wave cone of the
antiferromagnet. This is completely analogous to our above proposal
that the incoherent antinodal excitations are multiparticle states
composed of near nodal quasiparticles and incommensurate magnetic
excitations.\\
\begin{figure}
\centering
\includegraphics[width=5in,angle=0]{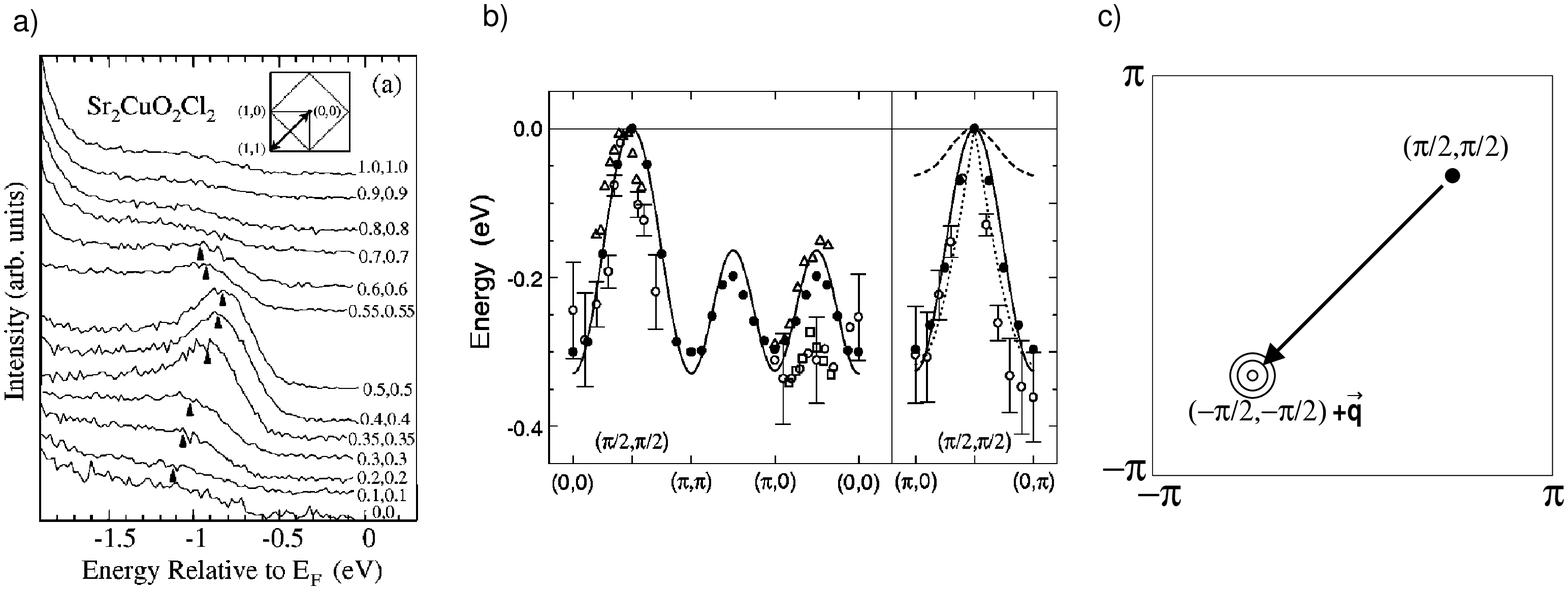}
\caption{ARPES spectra of insulating $Sr_2CuO_2Cl_2$, from
Damascelli
  et. al. \cite{zxreview}.  a) the broad feature
  corresponding to nodal excitations near $(\pi/2,\pi/2)$.  b) the dispersion of this feature along two directions. Experimental data
  points from Refs.\cite{scoc} are the open symbols.  The dispersion
  is isotropic around $(\pi/2,\pi/2)$.  c) The multiparticle state consisting
  of a spin wave with momentum $(-\pi,-\pi) + {\vec q}$ and a quasiparticle with momentum
  $(\pi/2,\pi/2)$ has the same quantum numbers as a photohole at
  momentum $(-\pi/2,-\pi/2) + {\vec q}$.
\label{singlehole}}
\end{figure}

In summary, we propose a mechanism for the decoherence of the
antinodal electronic excitations in the underdoped high temperature
superconductors. This mechanism attributes the broad antinodal
spectra seen in ARPES to the that of a multi-particle excitation
made up of a quasipaticle near the nodes and an incommensurate
antiferromagnetic excitations. This point of view is supported by our
renormalization group analysis.
\\

{\bf Acknowledgement}: We thank J.C. Davis, H. Ding, G.-H. Gweon, C.
Honerkamp, A. Lanzara, K. McElroy, K. Shen, Z.-X. Shen and X.-J.
Zhou for useful discussions. This work was supported by the
Directior, Office of Science, Office of Basic Energy Sciences,
Materials Sciences and Engineering Division, of the U.S. Department
of Energy under Contract No. DE-AC02-05CH11231 (DHL).

\widetext
\end{document}